\documentclass[review,12pt]{elsarticle}

\usepackage{amssymb}

\usepackage{graphicx}
\usepackage[colorlinks=true, allcolors=blue]{hyperref}
\usepackage{multicol}
\usepackage{caption}
\usepackage{subcaption}
\usepackage{tabularx}
\usepackage{float}

\journal{}

\begin{document}

\begin{frontmatter}



\title{Performance Evaluation of Pulse Shape Discrimination Capable Organic Scintillators for Space Applications}


\author[label1]{M.I. Pinilla-Orjuela\corref{cor}}
\ead{mpinilla@lanl.gov}
\author[label1]{K.E. Mesick}
\author[label1]{P.F. Bloser}
\author[label1]{J.R. Tutt}

\address[label1]{Los Alamos National Laboratory, Los Alamos, NM 87545 USA}
\cortext[cor1]{Corresponding author}

\begin{abstract}
Scintillators with pulse-shape discrimination (PSD) capability are of great interest to many fields in the scientific community. The ability to discern a gamma ray from a neutron using PSD varies between different types of scintillator materials and dopants. A new generation of organic scintillator materials with PSD capability were studied to determine their radiation hardness to ionizing and non-ionizing radiation. The PSD capability, average pulse shapes, and light output of four types of organic scintillator were characterized before and after neutron and gamma-ray irradiation. The main goal of this investigation is to study the effects of long-term irradiation that may be experienced in space applications on the light output and particle discriminating capabilities of each material. EJ-270, EJ-276, organic glass, and Stilbene were tested. Damage due to non-ionizing (neutron) radiation was not observed in any of the scintillators up to $2.56\times10^{11}$~n/cm$^2$, except for Stilbene which showed a small (12$\%$) decrease in light output. All scintillators presented some light output reduction after ionizing (gamma-ray) irradiation, with reductions of 17$\%$ (EJ-276 and OGS), 32$\%$ (EJ-270), and 42$\%$ (Stilbene) observed immediately after 100 kRad.

\end{abstract}




\end{frontmatter}


\section{Introduction}
\label{Intro}
The next generation of organic scintillators for fast neutron detection with pulse-shape discrimination (PSD) capability have recently been under development~\cite{Zaitseva15,Zaitseva12,Zaitseva13,Carlson}. These scintillators are of interest to a wide range of applications that benefit from fast neutron detection, including space-based applications such as planetary science and space science measurements. The PSD capability of these new organic scintillators provides a measure to cleanly reject gamma-ray background that organic scintillators are also sensitive to.

Some organic scintillators with PSD capability already exist. Stilbene is well known and has excellent PSD ability, however, until recently the availability of Stilbene has been limited and the cost of manufacturing large volumes high. A new growth method for Stilbene was recently developed at Lawrence Livermore National Laboratory (LLNL)~\cite{Zaitseva15}, which opens the door to easier scalability. In addition, Stilbene produced with the new growth method showed 50$\%$ more light output than Stilbene produced using the traditional growth method~\cite{Zaitseva15}. Liquid organic scintillators have also been used for decades and provide good PSD, however, the challenge of flying liquid scintillators and their required size on space missions makes them an unfavorable option. 

Recently, several new options for PSD capable organic scintillators have become available. In addition to the new Stilbene mentioned above, plastic scintillators with PSD capability, both unloaded \cite{Zaitseva12} and loaded with $^6$Li \cite{Zaitseva13} to provide thermal neutron sensitivity, and PSD glass~\cite{Carlson} have become available.

To our knowledge, none of these new organic scintillators with PSD capability have space heritage or have undergone all of the relevant environmental testing required to assess their ultimate performance for space-based applications. The required testing includes irradiation of the scintillators to assess their tolerance to damage from total ionizing dose (TID) and displacement damage effects, and characterizing their performance after radiation exposure to typical doses and fluences of particles experienced on orbit.

Instruments in inter-planetary space or in Earth orbit are subject to high fluences of energetic charged particles. In inter-planetary space and Earth orbits outside the radiation belts, instruments are subject to $\sim10^9$ protons/cm$^2$ over a 10 year mission lifetime from high-energy galactic cosmic rays (predominantly protons with an average energy of 100s of MeV). Solar energetic proton events can also result in an additional $\sim6\times10^{10}$ protons/cm$^2$ ($>$10 MeV protons) over the same duration. In low Earth orbit, instruments may additionally be subject to trapped protons in the radiation belts leading to higher proton flux, but these missions typically have a shorter duration. 

This research focuses on the effects of radiation damage on the performance of newly developed organic scintillators with PSD capability via neutron and gamma-ray irradiation, providing critical information for assessing the future use of these scintillators for space missions.

\section{Methods}
\label{Methods}
\begin{figure}
\centering
    \includegraphics[width=5.3in]{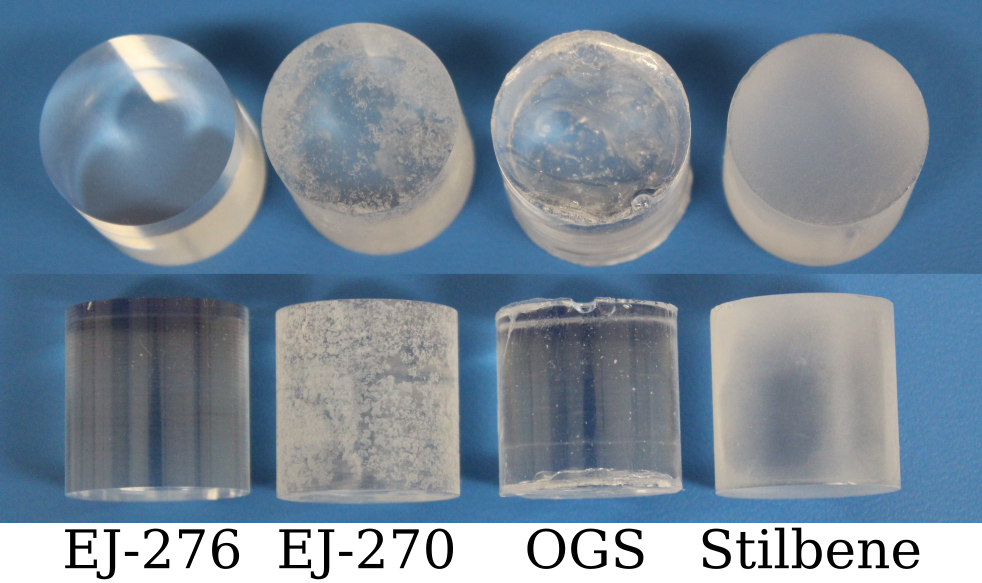}
    \caption{Picture of the four scintillator types obtained for this study.}
    \label{fig:samples}
\end{figure}

Seven 1" right cylinder samples were obtained of four different types of scintillators: unloaded PSD plastic (EJ-276, Eljen), $^{6}$Li-loaded PSD plastic (EJ-270, Eljen), Stilbene (InRad Optics), and organic glass scintillator (OGS, provided by Sandia National Laboratory).  The scintillator samples for each material numbered 1-7.  Figure~\ref{fig:samples} shows one sample of each scintillator type.  Sample 1 was measured five times over the duration of experimental measurements to establish a performance baseline and determine experimental uncertainty between measurements. Samples 2, 3, and 4 were irradiated with neutrons at the Los Alamos Neutron Science Center (LANSCE), while samples 5, 6, and 7 were irradiated with gamma rays at LANL's Radiation Instrument and Calibration Facility (RICF) Mark2b gamma cell irradiator. 

All scintillator samples were wrapped in four layers of polytetrafluoroethylene (PTFE) tape as uniformly as possible. They were then wrapped in electrical tape to maintain the integrity of the PTFE tape during handling and between measurements. The samples were coupled to 23~mm $\times$ 23~mm active area R11265U Hamamatsu photomultiplier tubes (PMTs) using optical grease. Waveforms were collected using a CAEN v1761 digitizer with a sample rate of 4~GSamples/s. The EJ-276 and EJ-270 samples were biased to $-700$V, the Stilbene to $-650$V, and the OGS to $-675$V, to limit input pulse amplitude to $<$1V as required by the digitizer.  

Each full set of characterization measurements consisted of 50,000 waveforms collected from $^{137}$Cs and $^{22}$Na check sources to obtain Compton Edge locations for energy calibration and 100,000 waveforms collected from a neutron source ($^{252}$Cf or PuBe).  Particle discrimination is achieved by using PSD, which is enabled by different scintillation light decay times for neutrons and gamma rays.  By integrating two regions of the scintillation light pulse, ``head" (H) and ``total" (T) regions, a PSD ratio is formed by $1 - H/T$.  The figure of merit (FOM) describes the quality of PSD, and is defined in Eq.~\ref{eq:FOM}~\cite{Knoll}:
\begin{equation}
    FOM = \frac{\mu_n - \mu_{\gamma}}{FWHM_n + FWHM_{\gamma}}~,
    \label{eq:FOM}
\end{equation}
where $\mu$ is the centroid of the neutron and gamma-ray peaks in PSD and FWHM their full-width and half-maximum.

\subsection{Neutron Irradiation}
\begin{figure}
\centering
    \includegraphics[width=5in]{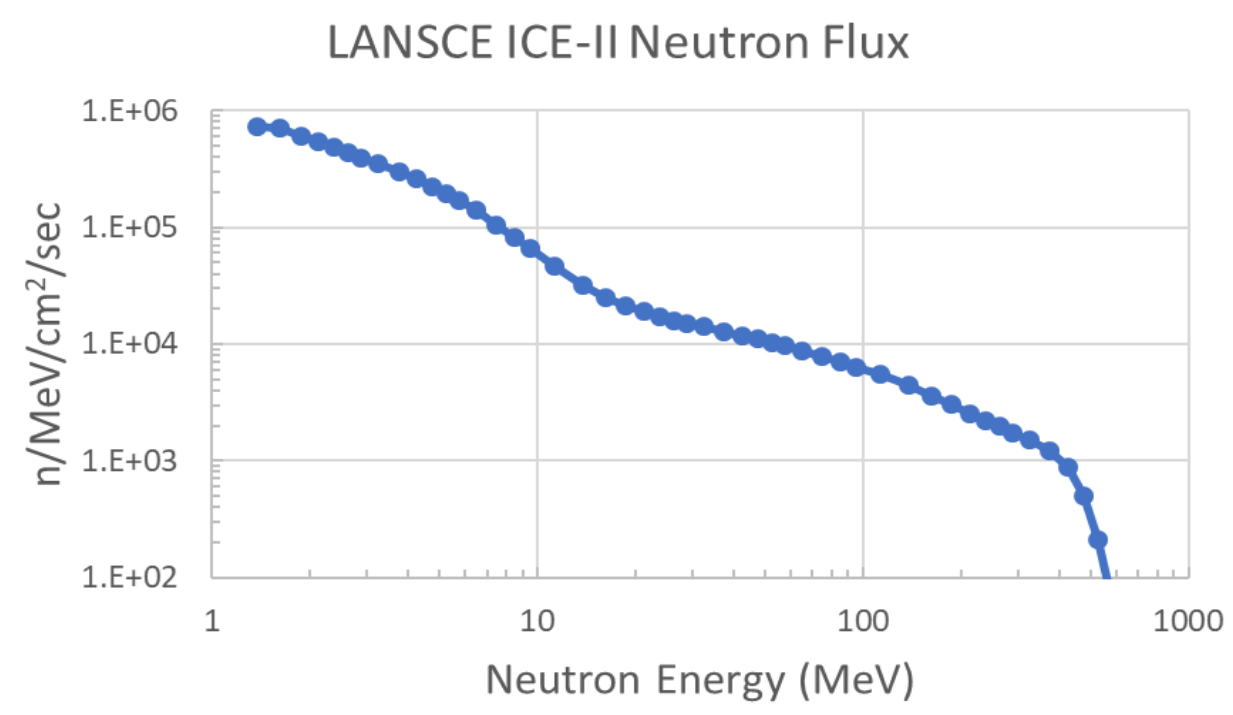}
    \caption{Neutron Spectrum for ICE-II flight path (30R) at LANSCE/WNR.}
    \label{fig:ICE2_flux}
\end{figure}

\begin{figure}
\begin{subfigure}{0.75\linewidth}
\centering
    \includegraphics[height=2.2in]{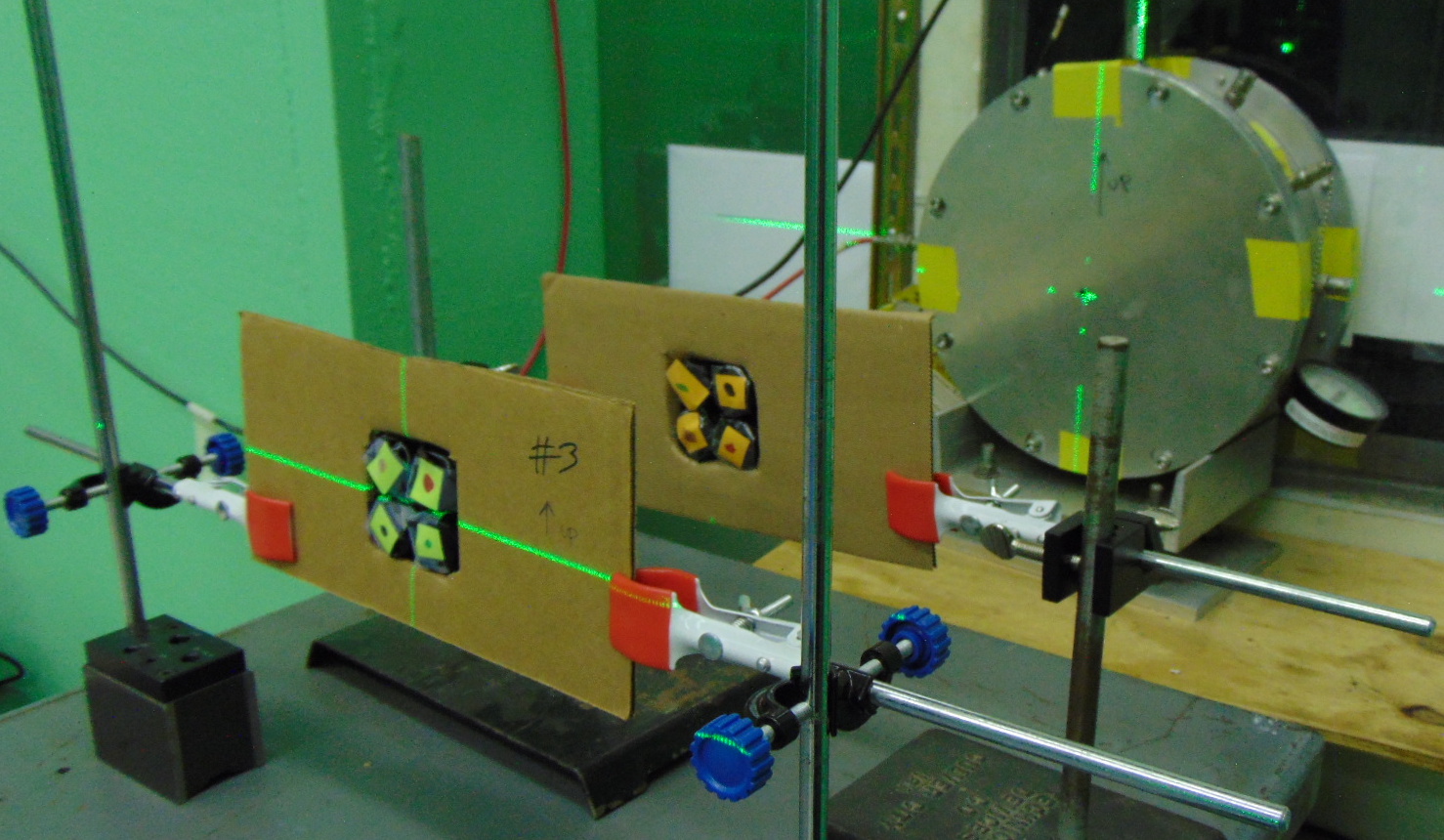}
    \caption{Scintillator arrangement during neutron irradiation.}
    \label{fig:LANSCEpic}
\end{subfigure}
\begin{subfigure}{0.24\linewidth}
\centering
    \includegraphics[height=2.2in]{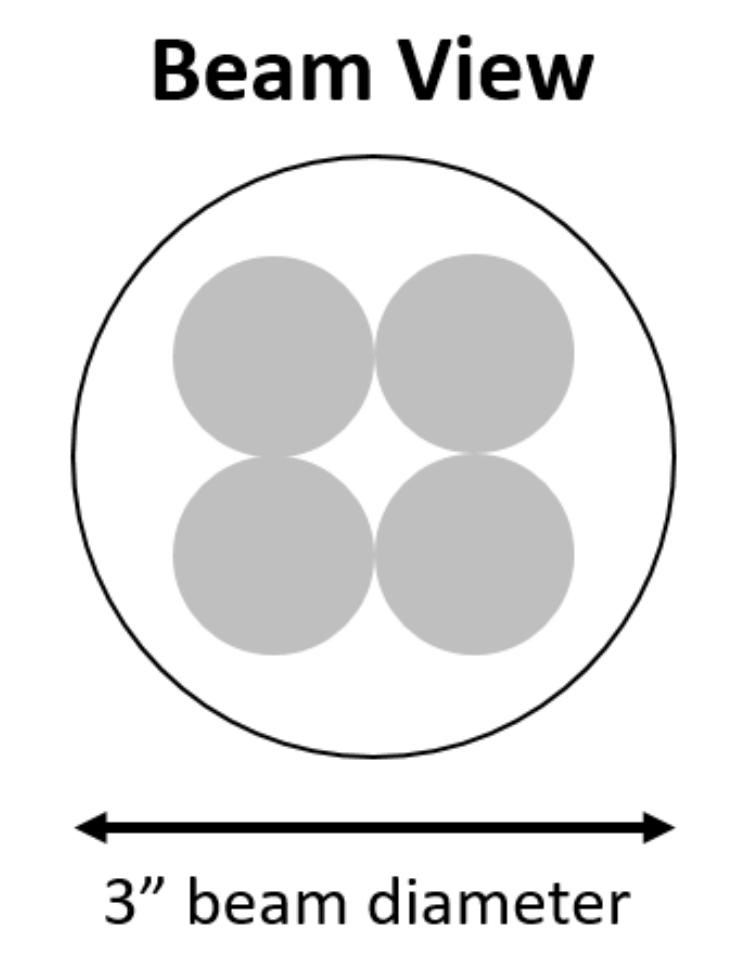}
    \caption{Beam view schematic.}
    \label{fig:beam_view}
\end{subfigure}
\caption{Experimental setup at LANSCE}
\label{fig:LANSCE}
\end{figure}

Neutron irradiation was used to study the effect of non-ionizing displacement damage on the scintillators. The Irradiation of Chips Electronics (ICE II) is located on the 30$^o$ flight path at the Weapons Neutron Research Facility (WNR) inside the LANSCE complex. The neutron beam at ICE II has an energy profile comparable to the neutron spectrum produced in the atmosphere by cosmic rays (see Fig.~\ref{fig:ICE2_flux})~\cite{Wender}. The high-intensity neutron flux allows for materials to be irradiated with high doses of radiation in a relatively short amount of time. Each scintillator sample set (EJ-270, EJ-276, OGS, and Stilbene) were placed along the beam path as seen in Fig.~\ref{fig:beam_view}. Sample sets 2, 3, and 4 were irradiated to achieve the neutron fluences seen in Table~\ref{tab:dose234} at an approximate rate of 1.8$\times$10$^6$~n/cm$^2$/s ($>$10~MeV).  The integral flux rate of $>$1~MeV neutrons is about two times higher.

\begin{table}
    \centering
    \begin{tabular}{c c c}
    \hline
        Sample & Time (hr) & $>$10~MeV Fluence (n/cm$^2$) \\
    \hline
        2    & 31.4 & 2.56$\times$10$^{11}$\\
        3    & 7.8 & 4.96$\times$10$^{10}$\\
        4    & 1.6 & 1.10$\times$10$^{10}$\\
    \hline
    \end{tabular}
    \caption{Irradiation times and neutron fluences ($>$10~MeV) achieved for scintillator samples 2, 3, and 4.}
    \label{tab:dose234}
\end{table}

\subsection{Gamma-ray Irradiation}
\begin{figure}
    \begin{subfigure}{0.5\linewidth}
    \centering
    \includegraphics[height=3in]{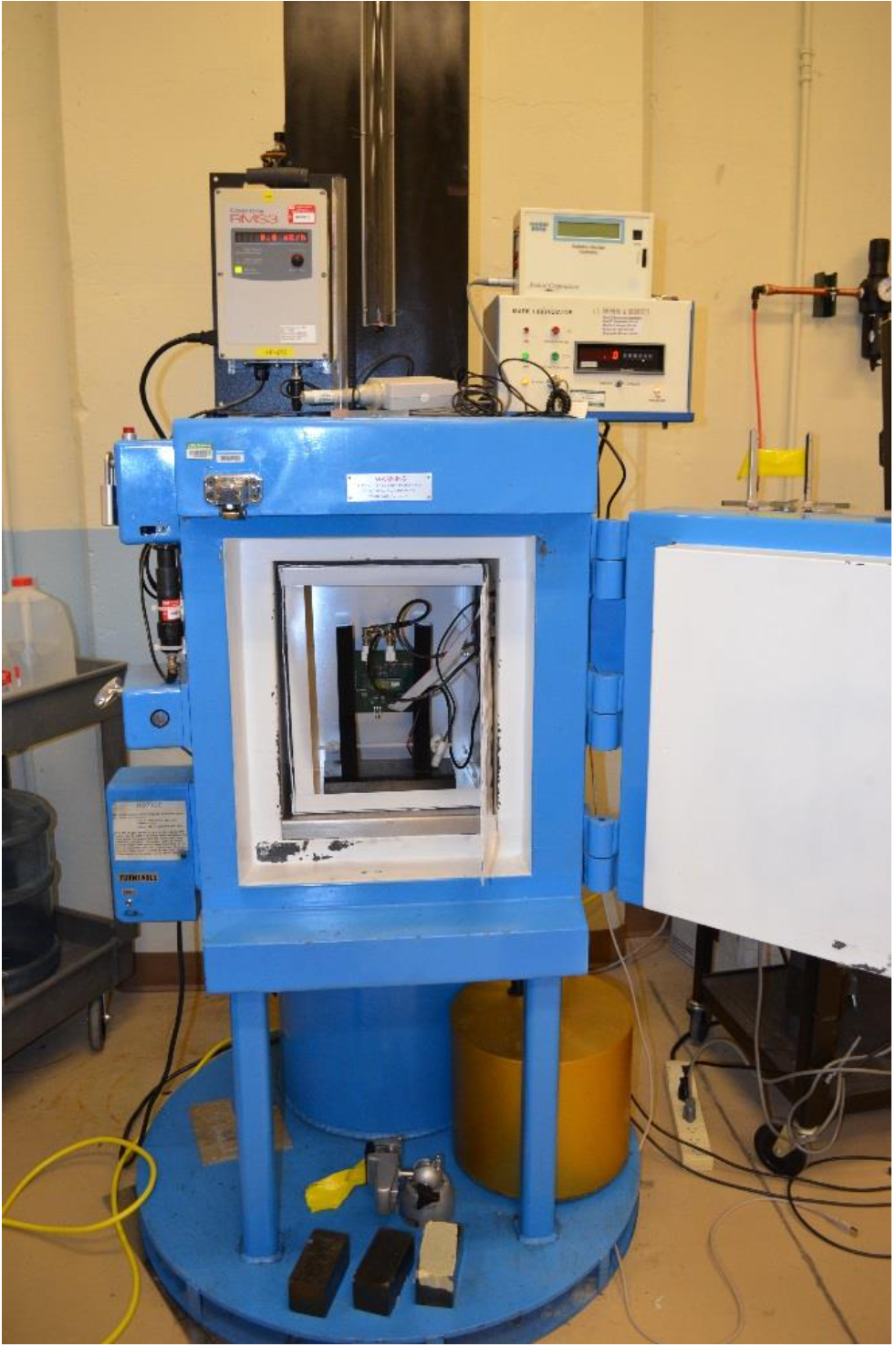}
    \caption{Mark2b Gamma Cell Irradiator}
    \label{fig:Mark2b}
  \end{subfigure}
  \begin{subfigure}{0.5\linewidth}
    \centering
    \includegraphics[height=3in]{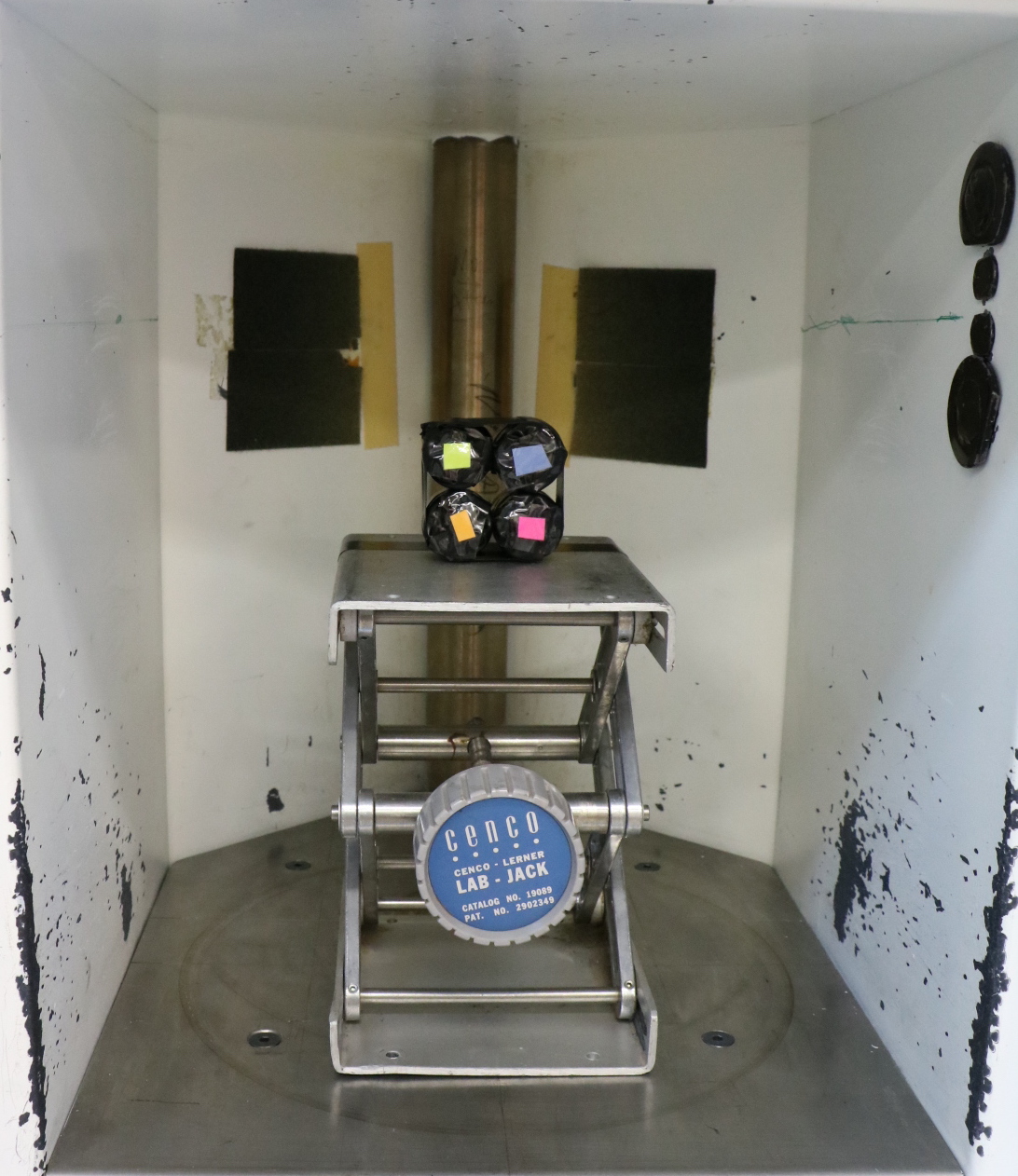}
    \caption{Scintillators inside Mark2b}
    \label{fig:inMark2b}
  \end{subfigure}
    \caption{Experimental setup at TA-36.}
    \label{fig:TA36}
\end{figure}

Scintillator samples 5, 6, and 7 were placed inside a Mark2b Gamma Cell Irradiator (see Fig.~\ref{fig:Mark2b}) containing a $^{137}$Cs source to assess the ionizing radiation hardness of the scintillators. Sample~5 received a dose of 100~kRad, sample~6 received 10~kRad, and sample~7 originally received 1~kRad; however, after seeing no effect with a 1~kRad dose, sample~7 was placed back in the chamber and received a total dose of 50~kRad.

\section{Results}
\label{Results}
\subsection{Initial Characterization}
Slight differences in manufacturing can change the light output and PSD capability of a detector. Therefore, we performed an initial characterization on all seven detector samples to understand the uniformity of their performance.  The Compton edge (CE) locations versus channel number for the 511~keV ($^{22}$Na, CE 340.67~keV), 667~keV ($^{137}$Cs, CE 477.34~keV), and 1.27~MeV ($^{22}$Na, CE 1061.71~keV) gamma-ray lines were compared to calculate the light output variance of the samples.  In addition, sample 1 of each scintillator type was measured five times to determine our measurement uncertainty.  The results for the four scintillator types are shown in Table~\ref{tab:LO_var}.  With the exception of OGS, the sample-to-sample light output variation was observed to be larger than our assessed measurement uncertainty.  
\begin{table}[h!]
    \centering
    \begin{tabular}{lcccc}
    \hline
         & EJ-270 & EJ-276 & OGS & Stilbene \\
    \hline
        Sample 1 (5 Meas.) & 3.79$\%$ & 3.69$\%$ & 3.66$\%$ & 3.26$\%$ \\
        Samples 1-7 & 7.5$\%$ & 5.0$\%$ & 3.4$\%$ & 13.2$\%$\\
    \hline
    \end{tabular}
    \caption{Sample light output variance and measurement uncertainty.}
    \label{tab:LO_var}
\end{table}

\begin{table}[h!]
    \centering
    \begin{tabular}{lcccc}
    \hline
         & EJ-270 & EJ-276 & OGS & Stilbene \\
    \hline
        Head (ns)  &  19 &  18 &  12 &  19 \\
        Total (ns) & 300 & 400 & 200 & 250 \\
    \hline
    \end{tabular}
    \caption{Integration windows used for PSD values.}
    \label{tab:Int_windows}
\end{table}

\begin{figure}[h!]  
  \begin{minipage}[b]{0.5\linewidth}
    \centering
    \includegraphics[width=\linewidth]{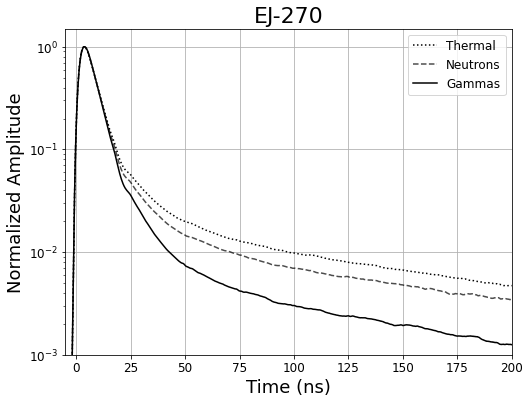}
  \end{minipage}
  \begin{minipage}[b]{0.5\linewidth}
    \centering
    \includegraphics[width=\linewidth]{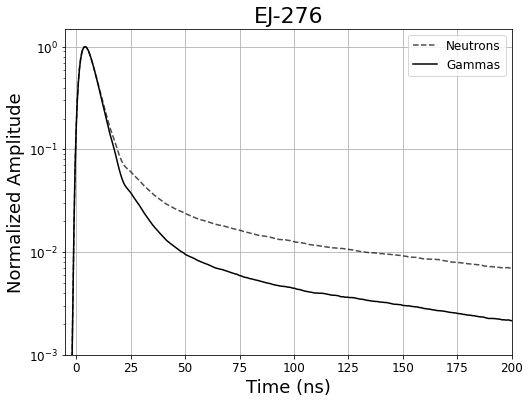} 
  \end{minipage} 
  \begin{minipage}[b]{0.5\linewidth}
    \centering
    \includegraphics[width=\linewidth]{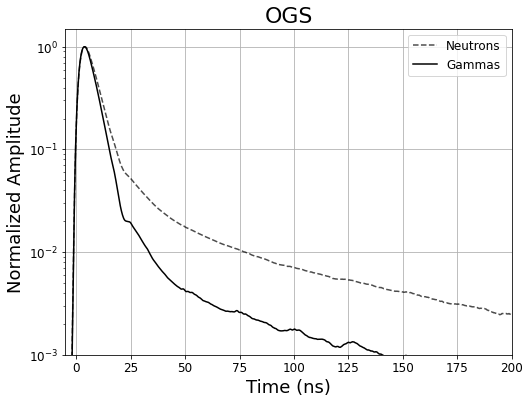} 
  \end{minipage}
  \begin{minipage}[b]{0.5\linewidth}
    \centering
    \includegraphics[width=\linewidth]{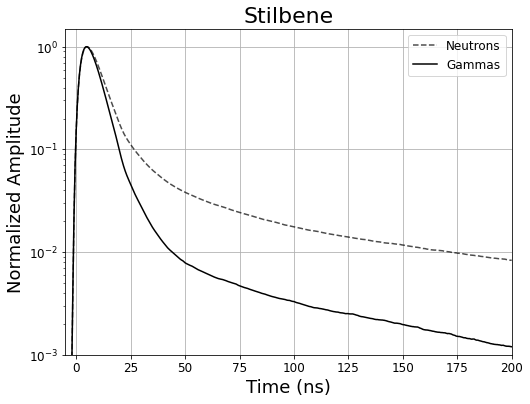} 
  \end{minipage} 
  \caption{Example average waveforms for each scintillator prior to irradiation.}
  \label{fig:avg_all4}
\end{figure}

\begin{figure}[h!]
  \begin{minipage}[b]{0.5\linewidth}
    \centering
    \includegraphics[width=\linewidth]{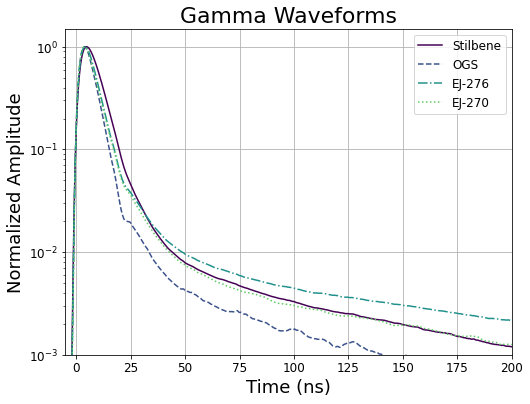} 
  \end{minipage}
  \begin{minipage}[b]{0.5\linewidth}
    \centering
    \includegraphics[width=\linewidth]{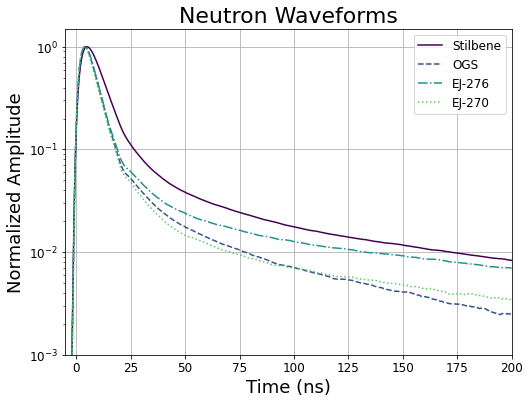} 
  \end{minipage}
    \caption{Comparison of average waveforms from the different scintillators.}
    \label{fig:avg_ng}
\end{figure}

\begin{figure}[h!]  
  \begin{minipage}[b]{0.5\linewidth}
    \centering
    \includegraphics[width=\linewidth]{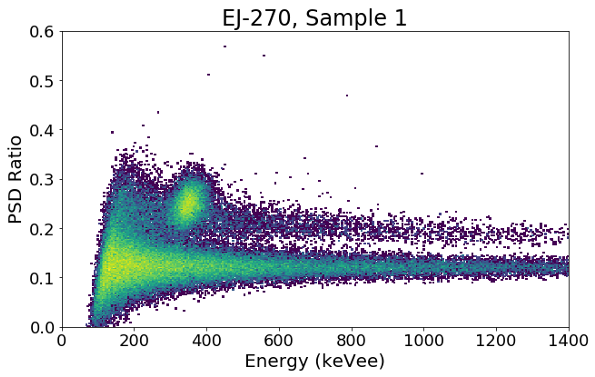}
  \end{minipage}
  \begin{minipage}[b]{0.5\linewidth}
    \centering
    \includegraphics[width=\linewidth]{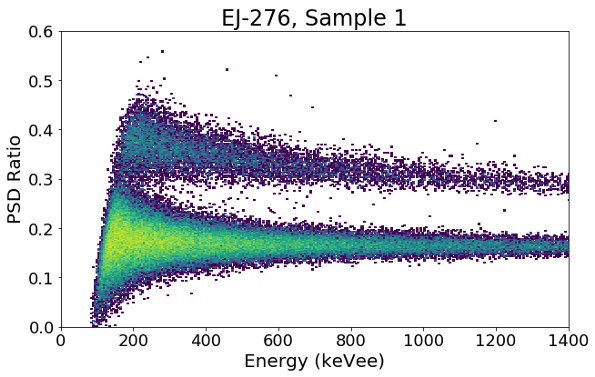}
  \end{minipage} 
  \begin{minipage}[b]{0.5\linewidth}
    \centering
    \includegraphics[width=\linewidth]{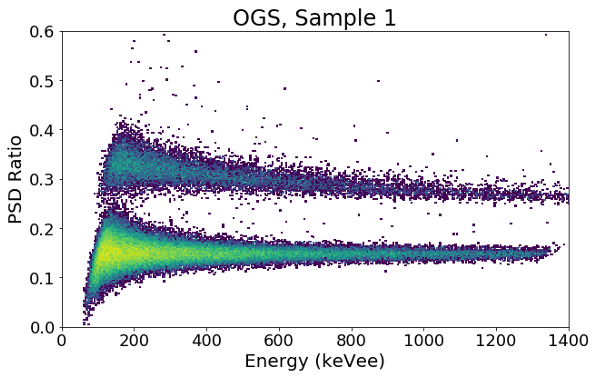} 
  \end{minipage}
  \begin{minipage}[b]{0.5\linewidth}
    \centering
    \includegraphics[width=\linewidth]{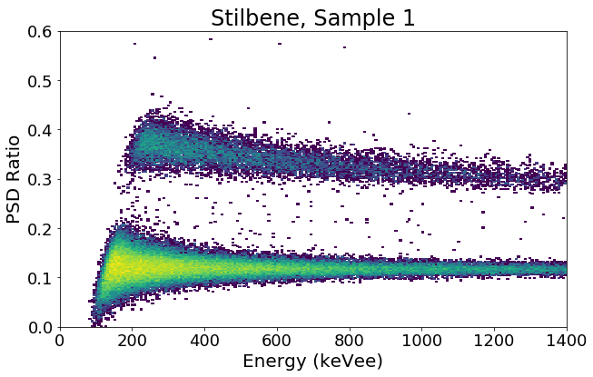} 
  \end{minipage} 
  \caption{Example PSD plots for each scintillator prior to irradiation.}
  \label{fig:PSDs}
\end{figure}

To test the uniformity of PSD performance among the samples, a $^{252}$Cf source was used to take combined neutron and gamma-ray data. Average gamma-ray and neutron waveforms for each detector are shown in Fig.~\ref{fig:avg_all4}. Figure~\ref{fig:avg_ng} shows the waveform comparison between the scintillators, with OGS showing the fastest decay for both neutrons and gamma rays. Due to the rapid decay of the gamma-ray waveform compared to the neutron waveform, we can calculate a PSD number based on the integral of the beginning of the waveform (head, H) to the total integral (T). The head and total integration windows, shown in Table~\ref{tab:Int_windows}, were optimized for each detector to maximize FOM (Eq.~\ref{eq:FOM}). This definition of a PSD value yields higher values for neutrons and lower values for gamma rays. Examples of the PSD versus calibrated electron-equivalent energy (ee) are shown for Sample 1 of each scintillator type in Fig.~\ref{fig:PSDs}. EJ-270 is sensitive to thermal neutrons, which can be seen in Fig.~\ref{fig:PSDs} as a “hot spot” between 200 and 400 keVee; the average thermal neutron waveform can be seen in Fig.~\ref{fig:avg_all4}.

\begin{figure}
    \centering
    \includegraphics[width=3.5in]{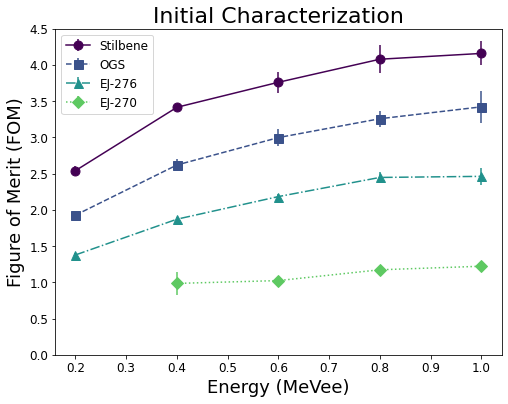}
    \caption{Figure of Merit - Initial characterization, averaged over 7 samples (lines for display purpose only)}
    \label{fig:FOM}
\end{figure}

Equation~\ref{eq:FOM} was then used to calculate the FOM for each detector every 200~keVee up to 1~MeVee; note that EJ-270 did not provide good enough separation to calculate a FOM value at 200~keVee and has some contamination from thermal neutrons at 400~keVee. The FOM and variance averaged over all seven samples of each detector at various energies can be seen in Fig.~\ref{fig:FOM} and listed in Table~\ref{tab:FOM_var}. The uncertainties in this Table include the measurement uncertainty obtained from the five repeated measurements of sample 1.  Stilbene has the highest FOM, followed by OGS, EJ-276, and EJ-270.

\begin{table}
    \centering
    \begin{tabular}{lcccc}
    \hline
    FOM & EJ-270 & EJ-276 & OGS & Stilbene \\
    \hline
    200 keVee &      N/A        & 1.37 $\pm$ 0.03 & 1.92 $\pm$ 0.03 & 2.54 $\pm$ 0.06  \\
    400 keVee & 0.95 $\pm$ 0.16 & 1.87 $\pm$ 0.01 & 2.60 $\pm$ 0.09 & 3.42 $\pm$ 0.04  \\
    600 keVee & 1.03 $\pm$ 0.04 & 2.18 $\pm$ 0.03 & 2.99 $\pm$ 0.10 & 3.80 $\pm$ 0.15  \\
    800 keVee & 1.17 $\pm$ 0.02 & 2.43 $\pm$ 0.08 & 3.24 $\pm$ 0.10 & 4.10 $\pm$ 0.18  \\
    1 MeVee   & 1.24 $\pm$ 0.04 & 2.48 $\pm$ 0.11 & 3.44 $\pm$ 0.19 & 4.23 $\pm$ 0.22 \\
    \hline
    \end{tabular}
    \caption{Average FOM and variance across the 7 samples.}
    \label{tab:FOM_var}
\end{table}

\subsection{Neutron Irradiation Effect}
\begin{figure}
    \centering
    \includegraphics[width=3.5in]{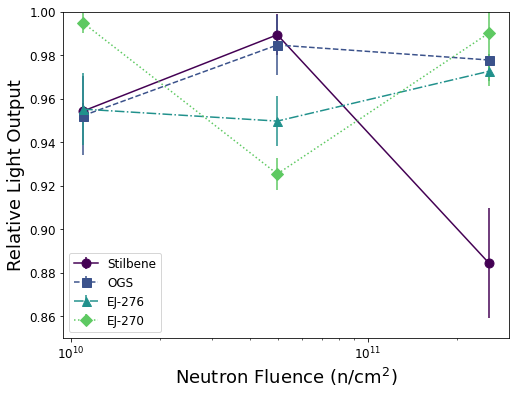}
    \caption{Light output after neutron irradiation, relative to pre-irradiation (lines for display purpose only).}
    \label{fig:LO_neutrons}
\end{figure}

\begin{figure}
    \centering
    \includegraphics[width=3.5in]{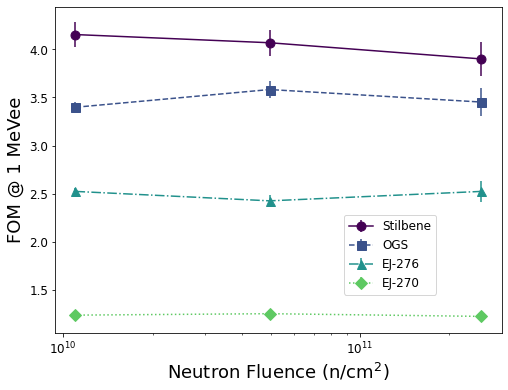}
    \caption{FOM after neutron irradiation (lines for display purpose only).}
    \label{fig:FOM_neutrons}
\end{figure}

After irradiating samples 2, 3, and 4 to the neutron fluences shown in Table~\ref{tab:dose234}, they were characterized one more time to determine whether there was any degradation in light output, average waveforms, or FOM due to non-ionizing radiation damage. Each of the samples was used to collect measurements using the $^{137}$Cs, $^{22}$Na, and PuBe sources. The location of the CE for the Cs and Na peaks were compared to their location in channel number prior to irradiation to quantify light output reduction. As seen in Fig.~\ref{fig:LO_neutrons}, there was no significant change in light output reduction except for Stilbene at the highest neutron fluence. The FOM at 1~MeVee is plotted against the neutron fluence received in Fig.~\ref{fig:FOM_neutrons}. Regardless of the neutron dose received, the average waveforms and FOM were not significantly affected for any of the samples. It is also important to note that there were no physical changes (e.g. yellowing) observed in any of the samples after the neutron irradiation.

\subsection{Gamma-ray Irradiation Effect}
\begin{figure}
     \centering
     \begin{subfigure}[b]{0.49\textwidth}
         \centering
         \includegraphics[width=2.5in]{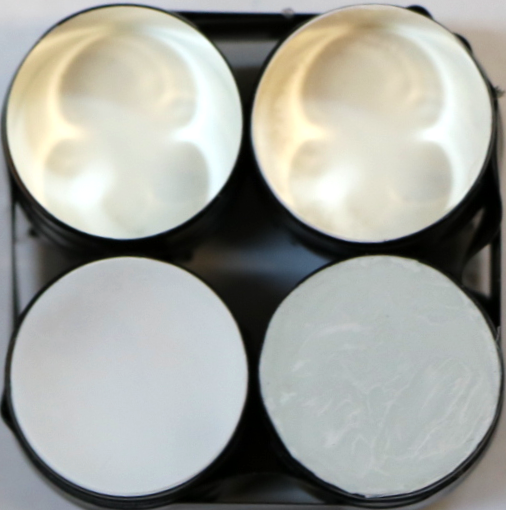}
         \caption{Before}
         \label{fig:before}
     \end{subfigure}
     \begin{subfigure}[b]{0.49\textwidth}
         \centering
         \includegraphics[width=2.5in]{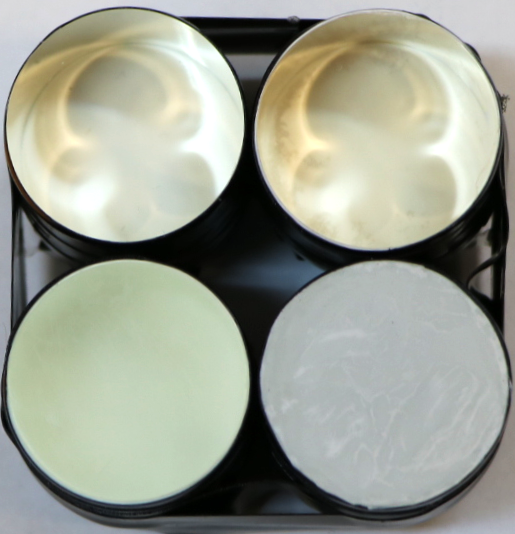}
         \caption{After}
         \label{fig:after}
     \end{subfigure}
        \caption{Scintillator samples before and after 100kRad irradiation; EJ-276 (top left), EJ-270 (top right), Stilbene (bottom left), OGS (bottom right).}
        \label{fig:before_after}
\end{figure}

Samples 5, 6, and 7 were exposed to  gamma-ray radiation using a $^{137}$Cs source. Sample 5 was exposed to a TID of 100~kRad, sample 6 to 50~kRad, and sample 7 to 1~kRad and 50~kRad. Stilbene and EJ-270 showed yellowing of the material after the 50 and 100~kRad exposures (see Fig.~\ref{fig:before_after}), with Stilbene having the most noticeable difference before and after irradiation. EJ-276 showed very little yellowing at the highest exposure, while OGS did not show any yellowing of the material. 

\begin{figure}
    \centering
    \includegraphics[width=3.5in]{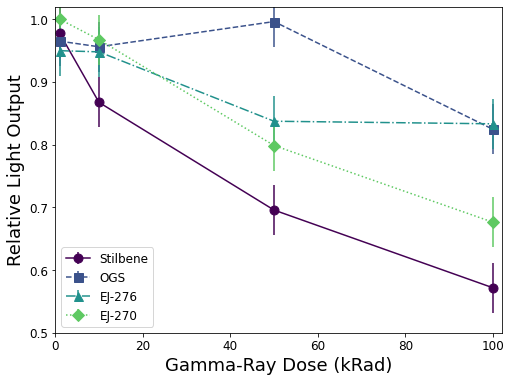}
    \caption{Light output after gamma-ray irradiation, relative to pre-irradiation (lines for display purpose only).}
    \label{fig:LO_gammas}
\end{figure}

Similar to the procedure after the neutron irradiation, the samples were characterized with $^{137}$Cs, $^{22}$Na, and $^{252}$Cf sources to determine the extent of radiation damage that had occurred. As expected from the qualitative observations of the materials post-irradiation, Stilbene presented with the highest reduction in light output at every exposure level as seen in Fig.~\ref{fig:LO_gammas}. EJ-270 also had a linear degradation in light output versus dose received, and was the second most damaged material. EJ-276 followed a similar light output reduction as EJ-270 up to 50~kRad, where the damage plateaued and no further reduction in light output was observed at the 100~kRad exposure level. In contrast, OGS showed no light output degradation below 50~kRad, but experienced similar damage as EJ-276 at 100~kRad of exposure. 

\begin{figure}
    \centering
    \includegraphics[width=3.5in]{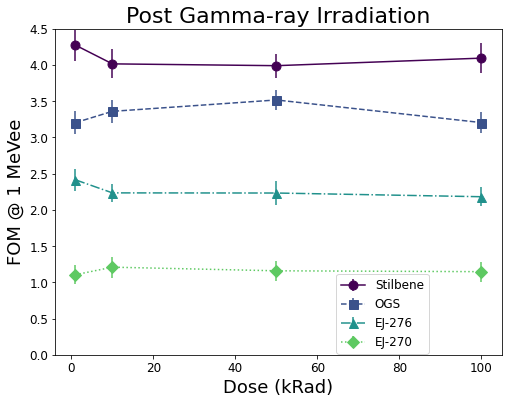}
    \caption{FOM after gamma-ray irradiation (lines for display purpose only).}
    \label{fig:FOM_gammas}
\end{figure}

The FOM of each sample was also calculated to determine whether the capability of each material to distinguish neutrons from gamma rays had been affected. The FOM at 1~MeVee versus dose received is shown  in Fig.~\ref{fig:FOM_gammas}. Regardless of exposure, all the samples retained their PSD capability, with the exception of Stilbene which had a slight decrease in FOM after 10 kRad of exposure.  The average waveforms were not significantly affected.

\begin{figure}
    \centering
    \includegraphics[width=3.5in]{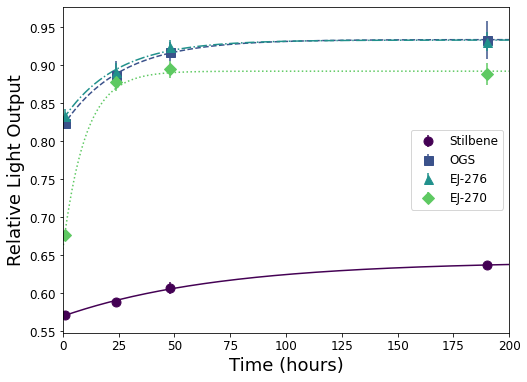}
    \caption{Annealing after 100 kRad gamma-ray irradiation.}
    \label{fig:100kRad}
\end{figure}

The samples that were exposed to 100~kRad were used to measure the gamma sources after 1~day, 2~days, and 1~week to determine whether the materials exhibited any annealing properties at room temperature. Stilbene showed very little improvement and slow recovery over time (time constant of 70~h), EJ-270 showed quick improvement in the first 24~hours with little to no recovery afterwards (time constant of 10~h), and EJ-276 and OGS (time constants of 24 and 26~h, respectively) showed similar improvement in light output when compared to their initial characterization (see Fig.~\ref{fig:100kRad}).  

\section{Conclusion}
\label{Conclusion}
The goal of this research was to characterize four organic scintillation detectors with PSD capability (EJ-270, EJ-276, OGS, and Stilbene), expose different samples of each material to varying doses of ionizing and non-ionizing radiation, and measure performance of the scintillators after radiation. Seven samples of each scintillation material were acquired. 
Samples 2, 3, and 4, were irradiated using neutrons to fluences of $1.10\times10^{10}$, $4.96\times10^{10}$, and $2.96\times10^{11}$ n/cm$^2$. Samples 5, 6, and 7 were exposed to gamma rays in a Mark2b Gamma Cell Irradiator to doses equivalent to 1, 10, 50, and 100~kRad. 

Samples 2, 3, and 4 were characterized after the neutron irradiation. No significant change was observed in light output reduction, average waveforms, or FOM for all the samples except for Stilbene, which showed marginal light output reduction and FOM degradation (5-7$\%$) at the highest neutron fluence. Post neutron irradiation, none of the samples showed any differences in coloration or visible damage.  

After the gamma-ray irradiation, samples 5, 6, and 7 were characterized to assess damage with total ionizing dose. Stilbene presented with yellowing of the material, highest light output degradation, and least recovery over time. EJ-270 showed yellowing of the material, second highest light output degradation, yet quick recovery. EJ-276 showed little yellowing, no additional damage $>$50~kRad, and quick recovery. OGS showed no yellowing, no damage $<$50~kRad, similar light output reduction to EJ-276 at 100~kRad, and similar recovery rate as EJ-276.  The average waveforms and FOM were not significantly affected, with the exception of Stilbene which showed a slight decrease in FOM after 10 kRad.

The decrease in light output observed in the experiments described is caused by radiation-induced damage in the scintillating materials. Ionizing radiation, such as gamma rays, give rise to color centers by displacing electrons which allow for new chemical bonds to form; Lima and Lameiras discuss this effect in the context of gemstones~\cite{Lima}. Color center formation also gives rise to absorption bands which reduce the light output of the scintillating material~\cite{Zhu}. The susceptibility of Stilbene to higher radiation damage after exposure to gamma rays is likely due to a combination of effects, including its crystalline structure, induced phosphorescence, and optical inhomogeneity~\cite{Gorelik}. In plastic scintillators, exposure to radiation can cause breaks and cross-linking of the polymer chains that make up the material, also giving rise to color centers which absorb scintillation light and ultimately reduce the light output of the material~\cite{Kharzheev}. The effect of radiation damage for different dose rates on plastic scintillators without PSD capability is also discussed in~\cite{Khachatryan}. Our experimental findings highly correlate previous literature, although dose-dependent radiation damage has not been previously compared between Stilbene, PSD capable plastic, and PSD capable glass scintillators. 

\section{Acknowledgements}
\label{Acknowledgements}
This work was supported by the U.S. Department of Energy through the Los Alamos National Laboratory and performed, in part, at the Los Alamos Neutron Science Center (LANSCE).  The authors would also like to acknowledge Patrick Feng and Lucas Nguyen of Sandia National Laboratory for providing the organic glass scintillator samples, Steve Wender, Kranti Gunthoti, and Jeff George for supporting the LANSCE measurements, and Charity Roybal, Nick Wehmann, and Dave Seagraves for supporting the gamma cell irradiation measurements.



\bibliographystyle{elsarticle-num} 
\bibliography{references.bib}

\end{document}